\documentclass[
    reprint, showpacs, showkeys,
 amsmath,amssymb,
 aps,
 pre
]{revtex4-2}

\usepackage{dcolumn}
\usepackage{bm}

\usepackage[all]{xy}\usepackage[latin1]{inputenc}        
\usepackage[dvips]{graphics,graphicx}
\usepackage{amsfonts,amssymb,amsmath,color,mathrsfs, amstext}
\usepackage{color}
\usepackage{amsbsy, amsopn, amscd, amsxtra, amsthm}
\usepackage{enumerate}
\usepackage{algpseudocode}
\usepackage{upref}
\usepackage{geometry}
\usepackage{relsize}
\usepackage{mathrsfs}
\usepackage{bm}
\usepackage{exscale}
\usepackage{graphicx}
\usepackage{tabularx}
\usepackage{threeparttable,multirow,dcolumn,booktabs}
\usepackage[linesnumbered,ruled,vlined]{algorithm2e}
\usepackage{tabularx}
\usepackage{caption}
\usepackage{float}
\usepackage{caption}
\usepackage{yhmath}
\usepackage{booktabs}
\usepackage{subcaption}
\usepackage{multirow}
\usepackage{makecell}
\usepackage[colorlinks,
            linkcolor=red,
            anchorcolor=red,
            citecolor=red
            ]{hyperref}
\usepackage{bm}
\usepackage[table]{xcolor}
\usepackage{booktabs}
\usepackage{verbatim}

\numberwithin{equation}{section}

\numberwithin{equation}{section}

\def\diag{\mathrm{diag}}

\definecolor{darkgreen}{RGB}{0,100,0}

\definecolor{orange}{cmyk}{0,0.5,1,0} 
\definecolor{purple}{RGB}{255,0,255}

\numberwithin{equation}{section}
\begin{document}

\preprint{APS/123-QED}

\title{Weighted balanced truncation method for approximating kernel functions by exponentials}

\author{Yuanshen Lin$^1$}
\author{Zhenli Xu$^1$}
\author{Yusu Zhang$^2$}
\author{Qi Zhou$^1$}
\email{zhouqi1729@sjtu.edu.cn}

\affiliation{$^1$ School of Mathematical Sciences, MOE-LSC and CMA-Shanghai, Shanghai Jiao Tong University, Shanghai, 200240, P. R. China} 
\affiliation{$^2$ Zhiyuan College, Shanghai Jiao Tong University, Shanghai, 200240, P. R. China}

\date{\today}

\begin{abstract}
Kernel approximation with exponentials is useful in many problems with convolution quadrature and particle interactions such as integral-differential equations, molecular dynamics and machine learning. This paper proposes a weighted balanced truncation to construct a modified model reduction method for compressing the number of exponentials in the sum-of-exponentials approximation of kernel functions. This method shows great promise in approximating long-range kernels, achieving over 4 digits of accuracy improvement for the Ewald-splitting and inverse power kernels in comparison with classical balanced truncation. Numerical results demonstrate its excellent performance and attractive features for practical applications.
\end{abstract}


\maketitle


\section{Introduction}
\label{sec:level1}

Approximating univariate kernel functions by exponentials is a useful technique for constructing fast algorithms of problems with convolution quadrature and particle interactions. 
The design of the so-called sum-of-exponentials (SOE) approximation has attracted broad interest in areas such as fast convolution \cite{lubich2002fast,greengard2018anisotropic},  electrostatic calculation \cite{gimbutas2020fast}, molecular dynamics simulation \cite{bellissima2017density,bellissima2020time,GAN2025Fast}, dynamics of magnetic nanoparticle \cite{taukulis2012coupled}, dynamics of non-Markovian systems \cite{wisniewski2024dynamic}, and DNA melting curves \cite{blossey2003reparametrizing}. Particularly, the kernel-independent SOE methods, including the black-box algorithm \cite{greengard2018anisotropic} and de la Vall\'ee-Poussin model reduction method \cite{gao2022kernel}, have been proposed, providing efficient tools for kernel summation problems.

The number of exponentials determines the processing efficiency of subsequent fast algorithms. The Laplace transform of the SOE results in a sum-of-poles (SOP) which has also a number of applications such as electromagnetics \cite{xu2013bootstrap}, nonreflecting boundary problems \cite{alpert2000rapid,jiang2004fast} , accelerating fast Gauss transform \cite{jiang2022approximating} and fast convolution transformation \cite{jiang2025dual}. In control theory, the SOP is the transfer function of linear dynamical systems, which can be compressed by building on the balanced truncation method and the square root method \cite{moore1981principal,antoulas2001approximation}. The model reduction (MR) technique of the balanced truncation plays a crucial role in further decreasing the number of exponentials, exhibiting a significantly faster convergence rate compared to other approaches such as the classical Prony's method \cite{hamming2012numerical}. However, the long-range nature of the kernel functions leads to the difficulty of efficient compression, and a direct use of the classical MR requires a large number of exponentials. Other methods such as the damping Newton method \cite{bremer2010nonlinear, gimbutas2020fast} and Remez algorithm \cite{barrar1970remez,hackbusch2019computation} are also applicable for SOE approximation, mostly for the $1/r$ Coulomb kernel.   

In this paper, we propose a weighted balanced truncation (WBT) method for improving the model order reduction of compressing the number of exponentials. By introducing weight functions into the balanced truncation process, the WBT enhances the uniformity of the approximation error distribution over a given interval. Numerical results show that the WBT achieves an improvement of over 4 digits of accuracy compared to the classical MR method \cite{gao2022kernel} for general long-range kernels. Moreover, for problems with near singularity at the end, the WBT method effectively captures local and global features, presenting a high efficient method in approximating these kernels such as Coulomb and inverse power functions. The WBT is a generalization of the classical balanced truncation method, which is a simple and efficient approach to construct an improved model reduction scheme for many problems such as high-dimensional dynamical systems \cite{burohman2023data,schafer2011dimension,sandberg2004balanced,ramirez2015application} and multiscale modelling \cite{fish2021mesoscopic,efendiev2012local}.

\section{Method}
\label{sec::SOE-Finite}

Given an error criteria $\epsilon$ and an $N$-term SOE series, the goal of this paper is 
to compress the number of exponentials such that $P$ is minimized under the error level:
\begin{equation}
    \label{eq::MR}
      \left| \sum\limits_{j=1}^{N}\omega_je^{-s_jr} - \sum\limits_{j=1}^{P}\widetilde{\omega}_j e^{-\widetilde{s}_j r}\right| < \epsilon, 
\end{equation}
for $r\in [0,M]$. In general, a preliminary and high-accurate SOE approximation of an interested kernel can be obtained by some kernel-independent techniques \cite{greengard2018anisotropic,gao2022kernel}. Minimizing the number of exponentials in a given interval will significantly improve the simulation efficiency. One option is to use the balanced truncation method following the work of \cite{moore1981principal,antoulas2001approximation}.
Here, we introduce a novel weighted balanced truncation method, which will promote the performance of compression, leading to an improved model reduction method.

To present the WBT idea, one starts from the Laplace transform of the $N$-term SOE series and represents the resultant SOP by
a matrix form, 
\begin{equation}
\label{eq::Laplace_Tras}
    \mathcal{L}\left[\sum_{j=1}^{N}\omega_je^{-s_jr}\right]=\sum_{j=1}^{N}\frac{\omega_j}{z+s_j}=\bm{c}(z\bm{I}-\bm{A})^{-1}\bm{b}, 
\end{equation}
where $\bm{A}=-\diag\{s_1,\cdots,s_N\}$ is an $N\times N$ diagonal matrix, $\bm{b}=(\sqrt{|\omega_1|},\cdots,\sqrt{|\omega_N|})^T$ and $\bm{c}=(sgn(\omega_1)\sqrt{|\omega_1|},\cdots,sgn(|\omega_N|)\sqrt{|\omega_N}|)$ are column and row vectors of dimension $N$, respectively. The sign function $sgn(\omega)=\omega/|\omega|$ for nonzero $\omega$ and $sgn(0)=0$. 

The matrix form of the SOP can be considered as the transfer function of the following linear dynamical system  
\begin{equation}
\label{eq::Linear system}
    \left\{
    \begin{array}{l}
    \bm{x}'(r)=\bm{A}\bm{x}(r)+w(-r)u(r)\bm{b},    \\
    y(r)=w(r)\bm{c}\bm{x}(r),
\end{array}
\right.
\end{equation}
where $u(r)$ and $y(r)$ are the input and output of this system, respectively, and $w(r)>0$ is a weight function. 
If $\hat{u}(z)$ and $\hat{y}(z)$ are Laplace transforms of the weighted input $w(-r)u(r)$ and output $y(r)/w(r)$, then they can be connected by the transfer function
\begin{equation}
\label{eq::Laplace_system}
    \hat{y}(z) = \bm{c}(z\bm{I}-\bm{A})^{-1}\bm{b} \hat{u}(z).
\end{equation}
Here we introduce the weight function $w(r)$ in order to construct a modified model reduction. In the case of the Heaviside function, i.e., 
 $w(r)=H(r)$ with $H(r) = 1$ for $r \ge 0$ and $0$ otherwise, it is applied in constructing the original balanced truncation method,
 and has been widely discussed \cite{moore1981principal,antoulas2001approximation}.

By the transfer function, the reduction on the SOE series can be performed by the explicit solution of the linear dynamical system \eqref{eq::Linear system},
\begin{equation}
    y(r)=\int_{-\infty}^{r}\bm{c}w(r)e^{\bm{A}(r-t)}\bm{b}w(-t)u(t)\mathrm{d}t.
\end{equation}
In this work, one assumes that the weight function $w(r)$ is compactly supported on the interval $[0,T]$.  
When $w(r)=1$ in this interval, it recovers the time-limited balanced truncation (TLBT) method \cite{goyal2019time,redmann2020lt2,gawronski1990model}. 
Define the solution operator $\mathcal{H}$ such that $y(r)=\mathcal{H}u(r)$, and $\mathcal{H}^{*}$ being the conjugate operator. Due to the compactness of the weight function, one can express them by,
\begin{equation}
\label{eq::operator}
\left\{
\begin{aligned}
    \mathcal{H}u(r)&=\int_{-\infty}^{+\infty}\bm{c}w(r)e^{\bm{A}(r-t)}\bm{b}w(-t)u(t)\mathrm{d}t\\
    \mathcal{H}^{*}y(r)&=\int_{-\infty}^{+\infty}\bm{b}^{*}w^*(-r)e^{\bm{A}^{*}(t-r)}\bm{c}w^{*}(t)y(t)\mathrm{d}t.
\end{aligned}
\right.
\end{equation} 
The key to reduce the linear system lies in calculating the singular values $\{\sigma_i\}$ of operator $\mathcal{H}$. Let these singular values be  
in an descending order with corresponding  eigenfunctions $\{u_i(r)\}$, i.e., one has $\mathcal{H}^{*}\mathcal{H}u_i(r)=\sigma_i^2u_i(r)$. Indeed, using Eq. \eqref{eq::operator}, one obtains the eigenfunction, 
 \begin{equation}
 \label{eq::eigen_eq}
   u_i(r)=\frac{1}{ \sigma_i^2}\bm{b}^{*}w^{*}(-r)e^{-\bm{A}^{*}r}\bm{Q}\bm{v},
 \end{equation}
 with
 \begin{equation}
 \label{eq::Q and v}
     \begin{aligned}
      \bm{Q}&= \int_{-\infty}^{+\infty}e^{\bm{A}^*t}\bm{c}^{*}\bm{c}e^{\bm{A}t}w(t)w ^{*}(t)\mathrm{d}t, \\
 \bm{v}&=\int_{-\infty}^{+\infty}e^{-\bm{A}t}\bm{b}w(-t)u_i(t)\mathrm{d}t.\\
\end{aligned}
 \end{equation}
 Substituting Eq.~\eqref{eq::eigen_eq} into the expression of $\bm{v}$ in Eq.~\eqref{eq::Q and v}, one has $ \sigma_i^2 \bm{v}=\bm{P}\bm{Q} \bm{v}$
with
\begin{equation}
\label{eq::P}
 \bm{P} = \int_{-\infty}^{+\infty}e^{\bm{A}t}\bm{b}\bm{b}^{*}e^{\bm{A}^*t}w(t)w^{*}(t)\mathrm{d}t.
\end{equation}
Such $\bm{P}$ and $\bm{Q}$ are usually called Gramians in control theory \cite{moore1981principal}. One finds that $\sigma_i^2$ is eigenvalue of matrix $\bm{P}\bm{Q}$, i.e. $\sigma_i=\sqrt{\lambda_i(\bm{P}\bm{Q})}$, where $\lambda_i(\bm{P}\bm{Q})$ denotes the $i$-th eigenvalue. One calculates the expressions in Eqs. \eqref{eq::Q and v} and \eqref{eq::P} to obtain the entries of matrix $\bm{P}$ and $\bm{Q}$,
\begin{equation}
    \label{eq::WBT_entry}
    \left\{
    \begin{array}{l}\bm{P}_{ij}=\sqrt{|\omega_i\omega_j|}I_w(s_i,\overline{s}_j)\\   
    
    \\
     
     \bm{Q}_{ij}=sgn(\overline{\omega}_i)sgn(\omega_j)\sqrt{|\omega_i\omega_j|}I_w(\overline{s}_i,s_j),
    \end{array}
    \right.
\end{equation}
where $\bar{s}$ denotes the conjugate of complex number $s$, and the weighted integral $I_w$ is defined by
\begin{equation}
    \label{eq::I_w}
    I_w(x,y):=\int_{-\infty}^{+\infty}e^{-(x+y)t}w(t)w^{*}(t)\mathrm{d}t.
\end{equation}

The WBT procedure starts by computing the Gramians $\bm{P}$ and $\bm{Q}$ using Eq.~\eqref{eq::WBT_entry}. One then performs Cholesky factorizations $\bm{P} = \bm{S}\bm{S}^*$ and $\bm{Q} = \bm{L}\bm{L}^*$, followed by the singular value decomposition $\bm{S}^*\bm{L} = \bm{U}\bm{\Sigma}\bm{V}^*$ with $\bm{\Sigma} = \mathrm{diag}\{\sigma_1, \sigma_2, \dots, \sigma_N\}$. Let $\bm{R} = \bm{S}\bm{U}\bm{\Sigma}^{-1/2}$. One takes the linear transform
$\bm{\widetilde{A}}=\bm{R}^{-1}\bm{A}\bm{R}$, 
$\widetilde{\bm{b}}=\bm{R}^{-1}\bm{b}$ and $\widetilde{\bm{c}}=\bm{c}\bm{R}$, together with the congruent transformations $\bm{\widetilde{P}} = \bm{R}^{-1} \bm{P} (\bm{R}^{-1})^{*}$ and $\bm{\widetilde{Q}} = \bm{R}^{*} \bm{Q} \bm{R}$. These two matrices become diagonal, 
$\bm{\widetilde{P}}=\bm{\widetilde{Q}}=\bm{\Sigma}.$
The singular values are arranged in descending order, enabling the extraction of the principal information here. Specifically, the $P \times P$ principal block $\bm{\widetilde{A}}_{P}$ of $\bm{\widetilde{A}}$ is extracted, and the first $P$ dimensions of the vectors $\bm{\widetilde{b}}$ and $\bm{\widetilde{c}}$ are selected to form new vectors $\widetilde{\bm{b}}_{P}$ and $\widetilde{\bm{c}}_{P}$. By the eigen-decomposition of $\widetilde{\bm{A}}_{P}$ such that  $\bm{\Lambda}=\bm{X}^{-1}\widetilde{\bm{A}}_{P}\bm{X}$ is diagonal, a new linear system $(\bm{\Lambda}, \hat{\bm{b}}, \hat{\bm{c}})$ is constructed with $\hat{\bm{b}}=\bm{X}^{-1}\widetilde{\bm{b}}_{P}$ and $\hat{\bm{c}}=\widetilde{\bm{c}}_{P}\bm{X}$. One then obtains a refined $P$-term SOE approximation of the original $N$-term SOE after the inverse Laplace transformation
\begin{equation}    \label{eq::inverse}
    \mathcal{L}^{-1}\left[\bm{\hat{c}}(z\bm{I}-\bm{\Lambda})^{-1}\hat{\bm{b}}\right]=\sum_{j=1}^{P}\widetilde{\omega}_je^{-\widetilde{s}_j r},
\end{equation}
where $-\widetilde{s}_j$ is the $j$th diagonal of $\bm{\Lambda}$ and $\widetilde{\omega}_j$ is the product of the $j$th identities of $\hat{\bm{b}}$ and $\hat{\bm{c}}$.

The Gramians $\bm{P}$ and $\bm{Q}$ in Eqs.~\eqref{eq::Q and v} and \eqref{eq::P} are positive definite, and thus their Cholesky factorizations exist. This leads to the validity of the WBT. The complete algorithm of the WBT method is summarized in Algorithm \ref{alg:example}.

\begin{widetext}
\begin{algorithm*}[H]
\caption{The weighted balanced truncation method}\label{alg:example}
\begin{algorithmic}[1]  
\Require For a given SOE with weight function $w(r)$ with $N$ exponentials, initialize the matrix and vectors  $\bm{A},\bm{b}$ and $\bm{c}$ by Eq.~\eqref{eq::Linear system}. Set the constant $P<N$. The algorithm is composed of the following steps.
\State  
Compute the Gramians $\bm{P}$ and $\bm{Q}$ by Eq.~\eqref{eq::WBT_entry}, perform Cholesky factorization for these two matrices such that $\bm{P} = \bm{S}\bm{S}^*$ and $\bm{Q} = \bm{L}\bm{L}^*$, and execute SVD factorization $\bm{S}^*\bm{L} = \bm{U}\bm{\Sigma}\bm{V}^*$ where  the diagonal matrix $\bm{\Sigma} = \mathrm{diag}\{\sigma_1, \sigma_2, \cdots, \sigma_{N}\}$. 
\State 
Set the transition matrix $\bm{R} = \bm{S}\bm{U}\bm{\Sigma}^{-1/2}$ to obtain the transformed linear dynamical system  $\widetilde{\bm{A}}=\bm{R}^{-1}\bm{A}\bm{R}, \widetilde{\bm{b}}=\bm{R}^{-1}\bm{b}$ and $\widetilde{\bm{c}} = \bm{c}\bm{R}$. Consequently, the resultant Gramians are diagonal, namely, $\widetilde{\bm{P}}=\widetilde{\bm{Q}}=\bm{\Sigma}$. 
\State
Extract the $P\times P$ principal block  of $\widetilde{\bm{A}}$, the first $P$ identities of $\widetilde{\bm{b}}$ and $\widetilde{\bm{c}}$, yielding $\widetilde{\bm{A}}_{P}$, $\widetilde{\bm{b}}_{P}$ and $\widetilde{\bm{c}}_{P}$. Perform eigen-decomposition $\widetilde{\bm{A}}_{P} = \bm{X}\bm{\Lambda}\bm{X}^{-1}$ such that 
$\bm{\Lambda}=-\diag\{\widetilde{s}_1, \cdots, \widetilde{s}_P\}$. Here, $\widetilde{s}_j$ is the exponent of the $j$th exponentials in the reduced SOE. Compute $\hat{\bm{b}} = \bm{X}^{-1}\widetilde{\bm{b}}_{P}$ and $\hat{\bm{c}} = \widetilde{\bm{c}}_{P}\bm{X}$. The weights are then calculated by $\widetilde{\omega}_j= \hat{\bm{b}}_{j}\hat{\bm{c}}_j,\, j = 1,2,\cdots,P$. 
\end{algorithmic}
\end{algorithm*}
 \end{widetext} 

It is noted that  the TLBT method is a special case of the WBT method, which has been applied to large scale systems \cite{kurschner2018balanced}, discrete-time systems \cite{duff2021numerical}, semi-Markovian jump systems based on generalized Gramians \cite{zhang2021time} and data assimilation \cite{konig2023time}, indicating that the WBT shall be also useful in many model order reduction problems besides the SOE approximation. 
One direct use is to construct sum-of-Gaussians (SOG) approximation to interacting and convolution kernels to design fast algorithms for particle systems \cite{greengard1991fast,predescu2020useries,greengard2018anisotropic,liang2023random,chen2025Yukawa} and nonlocal problems in
high-dimensional spaces \cite{Tausch2009FGT,exl2016accurate}. The optimal truncation $T$ and weight function $w(r)$ for specific problems require a systematic study and remain open issues.

\section{Results}
\label{sec::Num}

We illustrate the performance of the proposed WBT method with several numerical examples. Three benchmark examples are studied, including a smooth Ewald splitting kernel, the Coulomb kernel for different weights and the inverse power kernels, in comparison with results of the model reduction method with the  classical balanced truncation (denoted by `classical' in legends). Unless otherwise stated, all weight functions in the following results are truncated within their target intervals, and we emphasize that selecting weight function $w(r)=1$ in the WBT method is identical to the TLBT method, hence we will use these two descriptions interchangeably without further distinction in this section.

For the convenience of usage, we provide a comprehensive MATLAB software, VP-WBT \cite{VPWBT}, powered by the Multiprecision Computing Toolbox \cite{Multiprecision}. This software applies the Vall\'ee Poussin(VP)-sum method \cite{gao2022kernel} to generate an $N$-term high-precision SOE or SOG approximations for general kernels. And it reduces the series by the WBT method with customized weight functions. The following  numerical results can be simply reproduced through the visual interface of the software.

\subsection{Smooth Ewald-splitting kernel}
\label{sec::VPWBT}
 
Consider the Ewald splitting kernel $\mathrm{erf}(\Lambda r)/r$ with $\mathrm{erf}(\cdot) $ denoting the error function and $\Lambda $ being a positive constant. This kernel is often studied for Coulomb systems, resulted by the Ewald splitting of $1/r$ kernel. A large $\Lambda$ corresponds to a rapid decay of the kernel near the origin, making the SOE approximation more difficult. We consider the SOE approximation on interval $[0,10]$ with three parameters $\Lambda=10, 50$ and $100$. With SOE approximation of $N=500$ generated by the VP-sum method, the initial SOE series achieves the maximum errors  from $10^{-10}$ to $10^{-8}$ for the three cases.

\begin{figure}[ht!]
    \centering
    \includegraphics[width=0.5\textwidth]{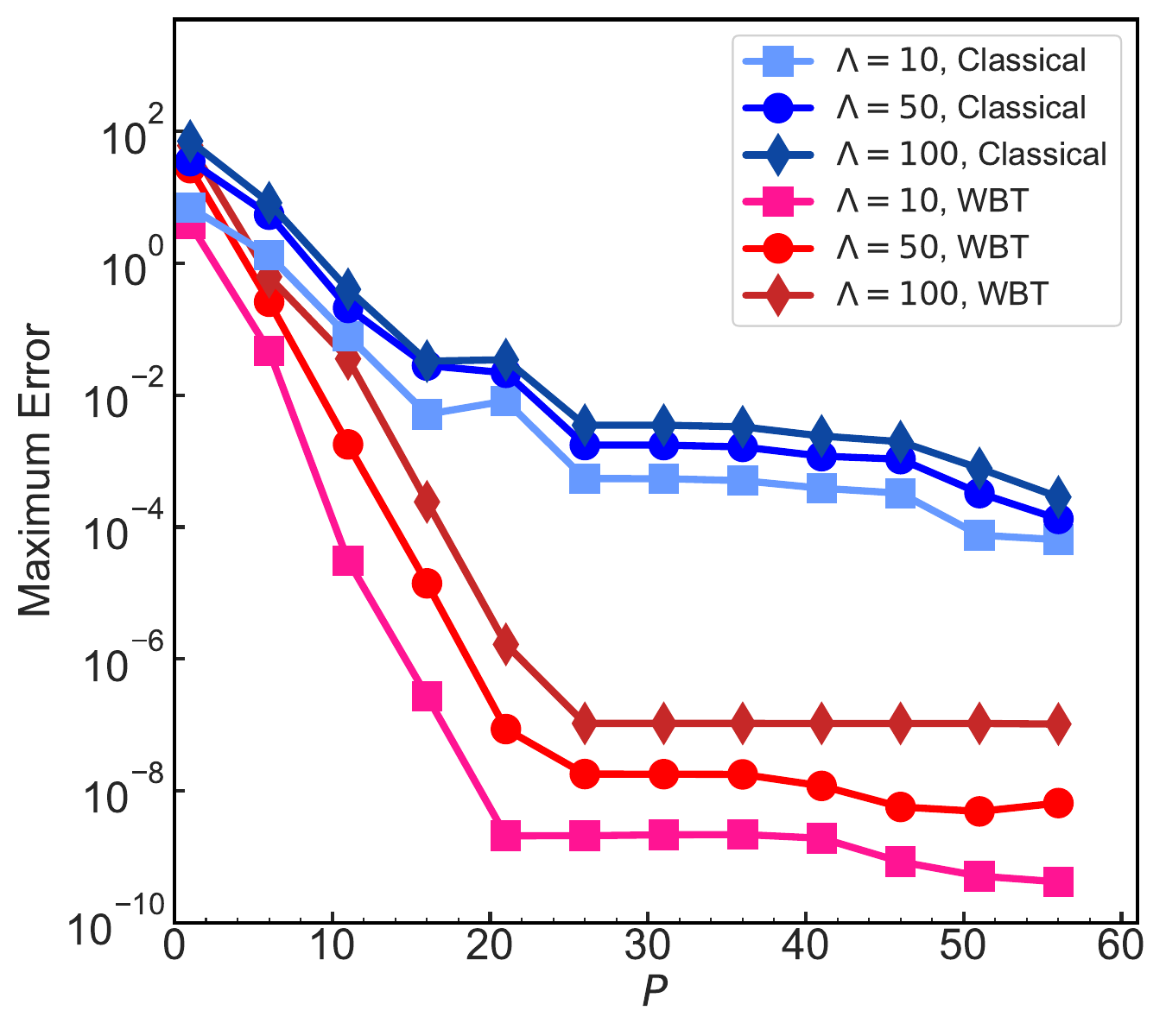}
    \caption{Maximum errors of SOE approximations of the Ewald splitting kernel on $[0,10]$ with respect to the reduction term $P$, computed using both the classical MR and the WBT with different Ewald splitting parameter $\Lambda=10,50$ and $100$.}
    \label{fig:ewald}
\end{figure}

In Fig.~\ref{fig:ewald}, we present the maximum errors of the WBT method and the classical MR results for the three $\Lambda$ with the increase of $P$. In the calculations, one sets the weight function $w(r)=1$ and truncation parameter $T=M=10$. One can observe the exponential decay of the error with $P$, rapidly approaching an error level under $10^{-8}$ with about 20 exponentials for  $\Lambda=10$.  It is noted that, a larger $\Lambda$ corresponds to a slower decay of the kernel, resulting in a slightly larger error. In the case of $P=20$ and $\Lambda=50,100$, the error is about $10^{-7}$ and $10^{-6}$.  In comparison, the classical MR method is at the level of $10^{-2}$ accuracy for $P=20$. For larger $P$, the WBT error remains nearly the same level as the original $500$-term SOE series. These results clearly demonstrate the rapid convergence of the WBT method in approximating smooth kernels.

\begin{figure*}[ht!]
    \centering
    \includegraphics[width=0.9\textwidth]{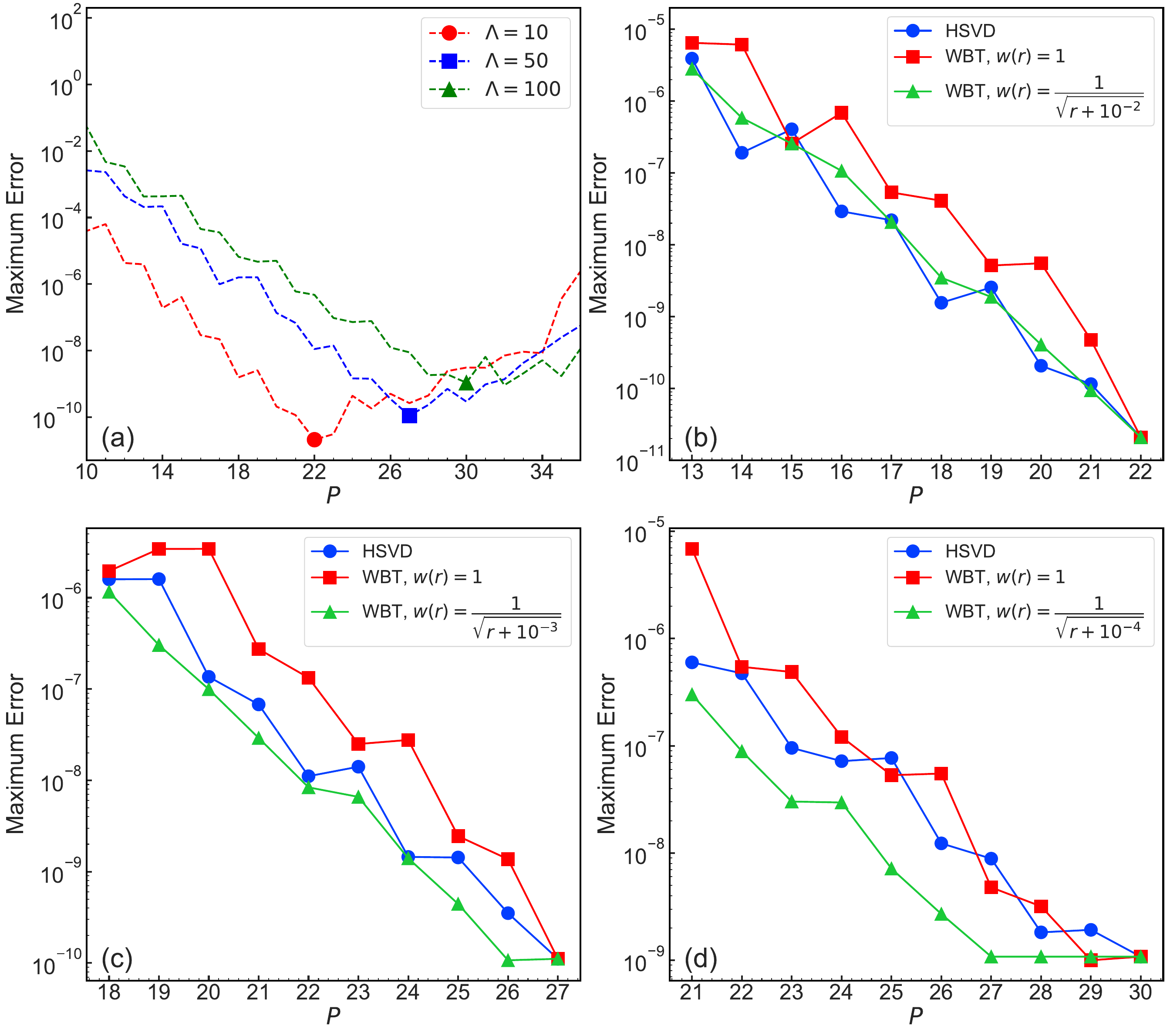}
    \caption{(a) The maximum errors in the SOE approximation using the HSVD method with respect to $P$, where the marked points indicate the highest achievable accuracy. 
(b-d) The compression accuracy versus the number of terms for the highest accuracy case using the HSVD and WBT methods.}
    \label{fig::ewald_2}
\end{figure*}

It is noted that the Hankel SVD (HSVD) method computes the Hankel singular values of dynamical systems, and can be also used to construct high-precision SOE approximation by exploiting the low-rank structure of Hankel matrices through truncated SVD  \cite{beylkin2019computing}. We perform the comparison between the WBT and the HSVD by using the same example as in Fig.~\ref{fig:ewald}. Fig.~\ref{fig::ewald_2}(a) presents the accuracy results obtained by the HSVD method with the increase of the exponential number $P$. One then starts from the initial SOE with the highest accuracy (correspondingly, $P=22, 27$ and $30$ for the three cases), and performs the compression by the WBT method.
Fig.~\ref{fig::ewald_2}(b-d) present the maximum errors of these reduction methods: the HSVD, the WBT with $w(r)=1$ (i.e., TLBT), and the WBT with a variable weight. The variable weights take $w(r)=1/\sqrt{r+d}$ with $d=10^{-2}, 10^{-3}$ and $10^{-4}$ for the three cases, respectively. For $\Lambda = 10$, the HSVD and the WBT with variable weight exhibits similar performance, better than the results reduced by the TLBT. For larger $\Lambda$, the WBT method demonstrates better performance than the HSVD. Particularly when $\Lambda = 100$, there is nearly one order of magnitude improvement by comparing the two methods. For all three cases, the TLBT method performs the worst compared to both the HSVD and the WBT methods. These results demonstrate the necessity of using weight functions in the balanced truncation, and the advantages of the WBT in dealing with long-range kernel functions.

\begin{figure}[htbp]
    \centering
    \includegraphics[width=0.5\textwidth]{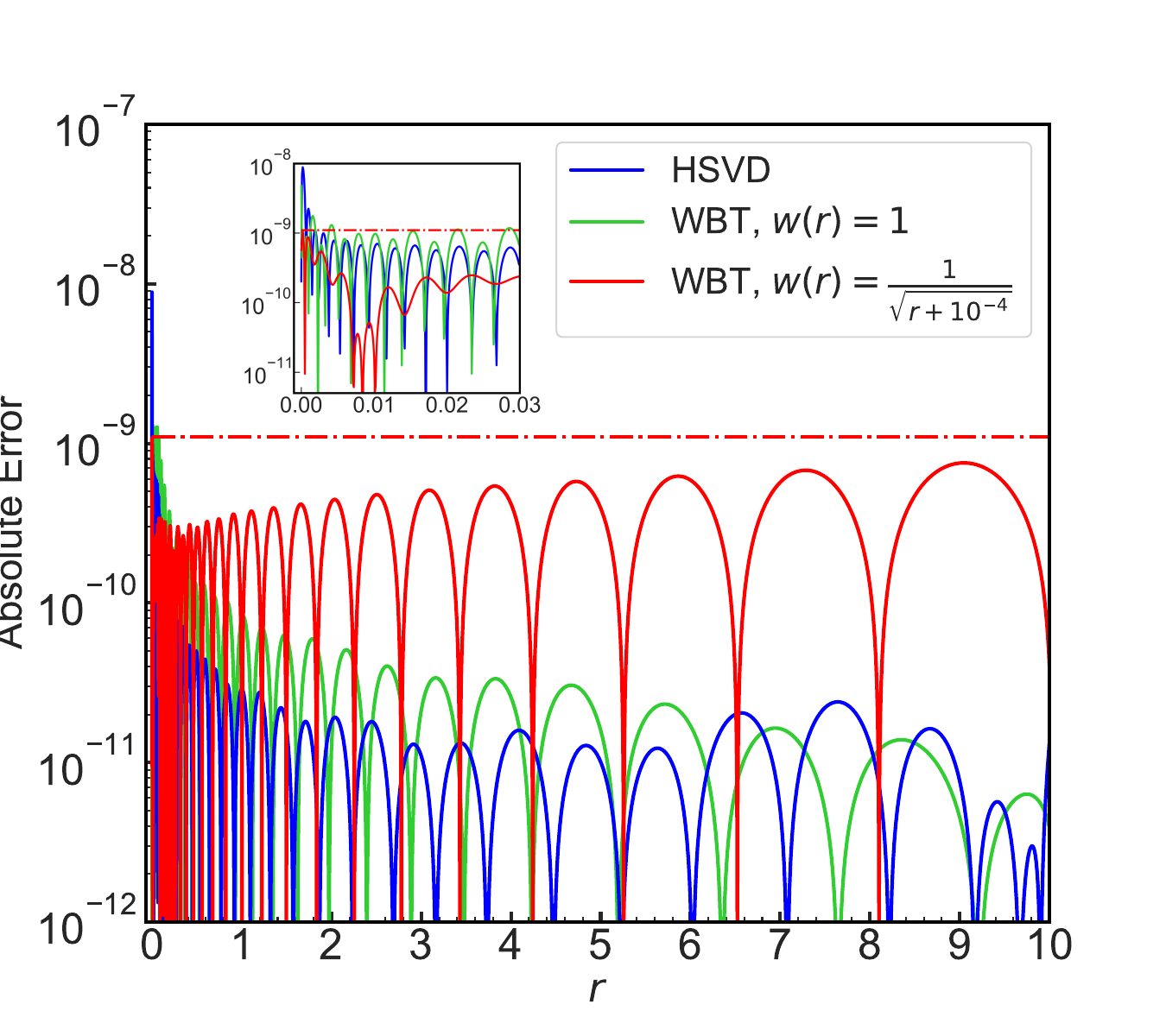}
    \caption{The error distribution of the SOE approximation to the Ewald splitting kernel at $ \Lambda = 100$, obtained by three different methods with $P=27$. The red dashed horizontal line represents the maximum error ($1.1\times 10^{-9}$) of the WBT with $w(r) = 1/\sqrt{r+10^{-4}}$.}
    \label{fig::ewald_27}
\end{figure}

For the most challenging case of $\Lambda = 100$, Fig.~\ref{fig::ewald_27} presents the error distribution over the interval $[0,10]$. One can observe that maximum error of the WBT with variable weight is about $10^{-9}$, while the maximum errors of the other two methods reach up to nearly $10^{-8}$. The use of a weight function leads to a more uniform error distribution, highlighting the crucial role of weight function in the approximation of long-range kernels. Similarly, for practical use, the coefficients and bandwidths of the 27-term SOE approximation are provided in Table~\ref{tab::Lambda100_ewald}. It is noted that, the exponents and weights of the preliminary SOE and during the model reduction for the HSVD and WBT methods can be complex numbers, revealing the broad applicability of the WBT method for different kinds of approximation by exponentials, e.g., in approximating oscillatory kernels.

\begin{table*}[ht!]
\centering
\caption{The SOE parameters for approximating Ewald-splitting kernel with $\Lambda=100$ by the WBT method with $P=27$. $\mathrm{Re}(\cdot)$ and $\mathrm{Im}(\cdot)$ represent the real and imaginary parts, respectively.}
\label{tab::Lambda100_ewald}
\begin{tabular}{cccc}
\toprule
$\mathrm{Re}(s_n)$ & $\mathrm{Im}(s_n)$ & $\mathrm{Re}(w_n)$ & $\mathrm{Im}(w_n)$ \\
\midrule
$351.021453049103$ & $\, 4.75818626608689\times10^{2}$ & $-0.178674220250501$  & $\, 1.60878709525225\times10^{-2}$ \\
$351.021452902964$ & $-4.75818626612809\times10^{2}$   & $-0.178674219581579$  & $-1.60878704192087\times10^{-2}$ \\
$348.444748992217$ & $\, 6.26669346953526\times10^{2}$ & $0.00184936871350844$ & $\, 1.40789958373811\times10^{-3}$ \\
$348.444748878602$ & $-6.26669346472470\times10^{2}$   & $0.00184936869572944$ & $-1.40789964418780\times10^{-3}$ \\

$345.422788497205$ & $\, 3.47679151746125\times10^{2}$ & $1.49747550227445$    & $-2.62594834524937\times10^{0}$  \\
$345.422788453179$ & $-3.47679151768061\times10^{2}$   & $1.49747550109614$    & $\, 2.62594834285696\times10^{0}$ \\
$329.956304037098$ & $\, 2.28042032083254\times10^{2}$ & $10.4901755839292$    & $\, 1.68434037779583\times10^{1}$ \\
$329.956304014370$ & $-2.28042032092664\times10^{2}$   & $10.4901755762687$    & $-1.68434037727680\times10^{1}$ \\
$296.901043744894$ & $\, 1.06998301350621\times10^{2}$ & $-66.5805514759647$   & $\, 7.75337350593186\times10^{0}$ \\
$296.901043734087$ & $-1.06998301362004\times10^{2}$   & $-66.5805514545991$   & $-7.75337351163996\times10^{0}$ \\
$175.982364054619$ & $-1.53074093064299\times10^{-9}$  & $82.6719475002774$    & $\, 3.13027338965423\times10^{-9}$ \\
$114.851016254158$ & $\, 3.30944687099560\times10^{-10}$& $46.1131765153415$   & $\, 3.73534114861676\times10^{-11}$ \\
$77.6366099298554$ & $-9.35531129907353\times10^{-11}$ & $29.9052333555613$    & $\, 9.91980314462552\times10^{-11}$ \\
$52.9356172959511$ & $\, 4.67272019388614\times10^{-10}$& $20.2208603064554$  & $\, 4.06460287252393\times10^{-10}$ \\
$36.1021227337447$ & $\, 1.96971020688910\times10^{-10}$& $13.8648027067850$  & $-5.61671049714930\times10^{-10}$ \\
$24.5356248845662$ & $\, 7.41256042732792\times10^{-11}$& $9.53775666107750$  & $\, 3.11584753997664\times10^{-10}$ \\
$16.5828186546525$ & $\, 2.11968606122945\times10^{-10}$& $6.55030820518430$  & $-1.52457361399390\times10^{-10}$ \\
$11.1308271185660$ & $\, 5.00765620859347\times10^{-11}$& $4.47994857927913$  & $-6.44821960349536\times10^{-11}$ \\
$7.41130471835977$ & $\, 4.07531588565541\times10^{-11}$& $3.04727271524753$  & $-8.29864308292231\times10^{-12}$ \\
$4.88825376924659$ & $\, 2.23947715694301\times10^{-11}$& $2.06058207682297$  & $-1.34559962001135\times10^{-11}$ \\
$3.18637922211194$ & $\, 1.01064614505056\times10^{-11}$& $1.38651068711969$  & $-1.20521027166503\times10^{-11}$ \\
$2.04226820614922$ & $\, 6.07192581564823\times10^{-12}$& $0.932236404822627$ & $-1.05532431704878\times10^{-12}$ \\
$1.27075281230222$ & $\, 1.62648642985215\times10^{-12}$& $0.631849303083841$ & $-3.90387788553250\times10^{-12}$ \\
$0.744786509078985$& $\, 5.08788513972153\times10^{-13}$& $0.432995938615514$ & $-8.72740533466332\times10^{-14}$ \\
$0.385901750556541$& $\, 1.04444276395237\times10^{-13}$& $0.291247508916668$ & $-3.56366935349646\times10^{-13}$ \\
$0.153444685696070$ & $\, 1.87588392500376\times10^{-14}$& $0.176579557326030$& $\, 3.46973306922824\times10^{-15}$ \\
$0.0287803364908740$& $\, 2.06424931760326\times10^{-15}$& $0.0740591567263405$& $-1.79511593057375\times10^{-14}$ \\

\bottomrule
\end{tabular}
\end{table*}

\subsection{Inverse power kernel}
\label{sec::BSAWBT}
Consider the inverse power kernel $f(r)=r^{-\alpha}$. It has a preliminary SOE series by the bilateral series approximation (BSA) \cite{beylkin2005approximation,beylkin2010approximation},  
 \begin{equation}
    \label{eq::BSA_formula}
    r^{-\alpha}\approx \frac{\sigma\log(b)}{\Gamma ( \alpha)}\sum_{\ell=-\infty}^{+\infty}b^{\alpha \ell}e^{-b^\ell \sigma r},
\end{equation}
where $\sigma$ represents a scaling factor of the bandwidth, and $\Gamma(\cdot)$ denotes the gamma function. The base parameter $b > 1$ determines the accuracy of the BSA approximation, which converges rapidly as $b$ asymptotically approaches 1.

\begin{figure}[htbp]
    \centering
    \includegraphics[width=0.5\textwidth]{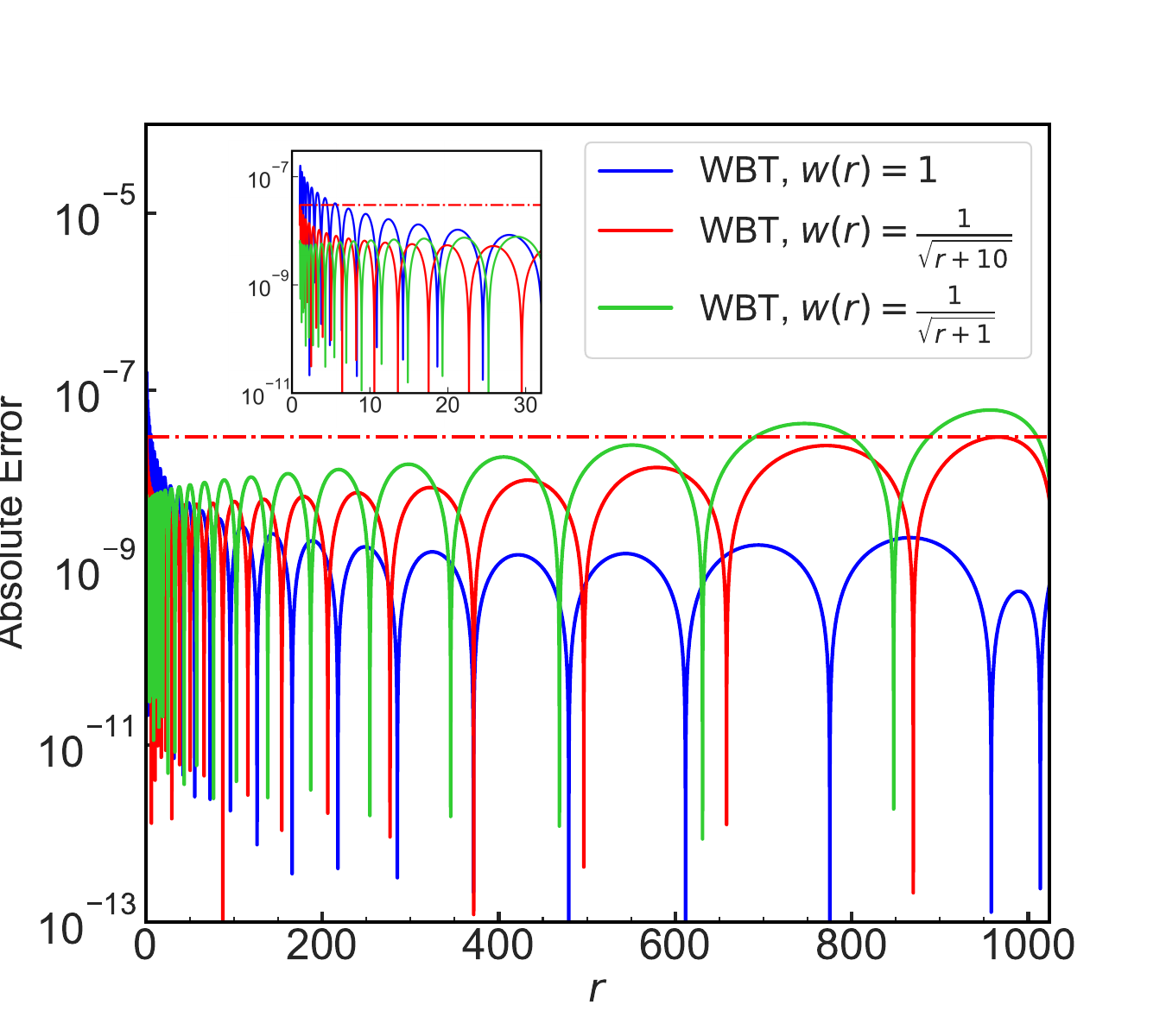}
    \caption{The error distribution of the WBT for the Coulomb kernel with different weight functions. The approximation interval is $[1,1024]$ with $P=15$. The red dashed horizontal line represents the maximum error ($3.0\times 10^{-8}$) of $w(r) = 1/\sqrt{r+10}$.}
    \label{fig:weighted}
\end{figure}

One first investigates the influence of the weight function $w(r)$ for the WBT method.  One considers the case of $\alpha=1$, i.e. the Coulomb kernel $1/r$ on $[1,1024]$ using three different weight functions $w(r)= 1$, $1/\sqrt{r+1}$ and $1/\sqrt{r+10}$. One takes $\sigma=1$ and $b=1.1$ in the BSA to obtain the preliminary SOE and sets the truncation parameter $T=512$ in the WBT.
Fig.~\ref{fig:weighted} presents the error distributions with $P=15$ for the three weights. One can observe that the error distribution for $w(r)=1$ is quite nonuniform. The error near  $r=1$ is much larger than the region away from the origin. The error distributions with the other two weight functions behave much better. Among the three weights, the maximum error of the $w(r) = 1/\sqrt{r+10}$ case is the smallest, which is $3.0\times 10^{-8}$. For comparison, the maximum errors for the  $w(r)\equiv 1$ and $1/\sqrt{r+1}$ cases are $1.9\times 10^{-7}$ and $6.0\times 10^{-8}$, respectively. The results demonstrate that the WBT can be very efficient when an appropriate weight function is employed.

The second experiment considers high accuracy approximation by the BSA in Eq.~\eqref{eq::BSA_formula} for the  preliminary SOE approximation of $N=500$ on the interval $[1, 1024]$ with $\sigma=1$ and $b=1.1$, which is at the machine precision. In the WBT, one selects $T=512$ and weight function $w(r) = 1/\sqrt{r+10}$. 
\begin{figure*}[t!]
        \centering
        \includegraphics[width=0.9\textwidth]{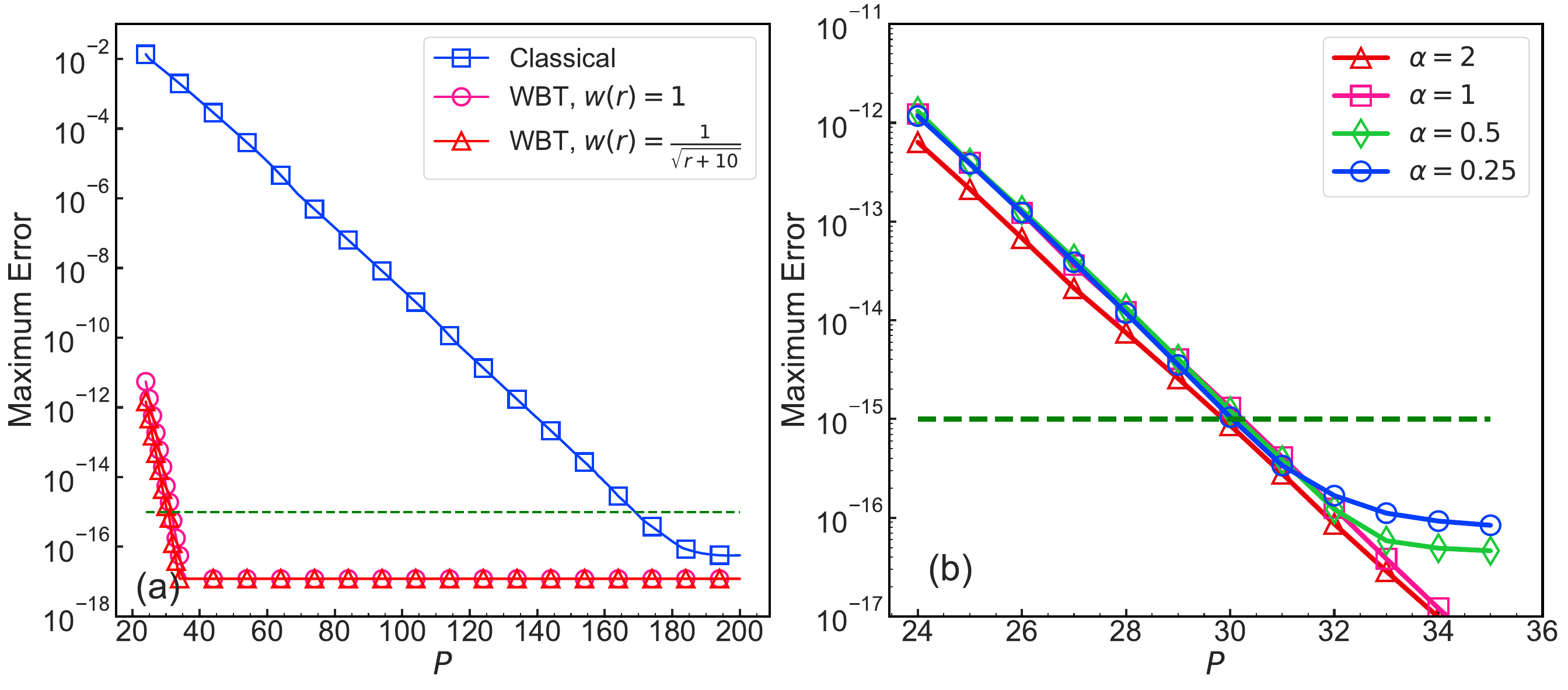}
    \caption{Maximum errors of SOE approximations of power function kernel on $[1,1024]$ with respect to the reduction term $P$. (a) Coulomb kernel reduced by different model reduction techniques, (b) Singular power functions with different $\alpha$ reduced by the WBT. The green dashed  line represents the precision of $1.0\times 10^{-15}$. }
    \label{fig:experiment1}
\end{figure*}
We first examine the accuracy of the Coulomb kernel for the $\alpha = 1$ case with varying $P$, and the results are present in Fig.~\ref{fig:experiment1}(a). One observes that the classical MR method exhibits a slow convergence rate in reducing the BSA sequence, while the WBT demonstrates remarkably fast convergence with over $9$ digits of accuracy improvement for $P>24$, achieving the maximum error of $7\times 10^{-16}$ with $31$ terms. This significant improvement arises because the WBT avoids the influence of long-range contributions outside the interval, which could otherwise affect the reduction and extraction of principal information. In contrast, the classical MR method has limitations in this regard, leading to slow convergence and low limit accuracy. Compared to the VP-sum with equidistant bandwidths in Section \ref{sec::VPWBT}, this advantage of the WBT method is particularly evident when applied to the BSA with exponentially distributed bandwidths. Indeed, the approximation provided by the WBT has similar performance as the well-known results reported by Gimbutas \emph{et al.} \cite{gimbutas2020fast} and Hackbusch \emph{et al.} \cite{hackbusch2019computation}. Remarkably, though the WBT method is a general-purpose model order reduction technique, it can still achieve accuracy comparable to methods designed for specific kernel functions, demonstrating its promising for broader applications. Moreover,
even better results can be achieved by leveraging optimization techniques with  detailed analysis of the weight function $w(r)$ and truncation parameter $T$. 

For different inverse power kernels, Fig.~\ref{fig:experiment1}(b) presents the convergence results with $\alpha = 0.25, 0.5, 1$ and $2$. With the same approximation interval and accuracy requirements, the WBT method delivers highly consistent approximation performance with the case of $\alpha=1$. It achieves the precision of $1.0\times 10^{-15}$ for all the cases for $P=31$, demonstrating that the WBT can effectively achieve an attractive behavior of SOE approximation for various forms of error decay tails.

\begin{figure}[htbp]
    \centering  \includegraphics[width=0.5\textwidth]{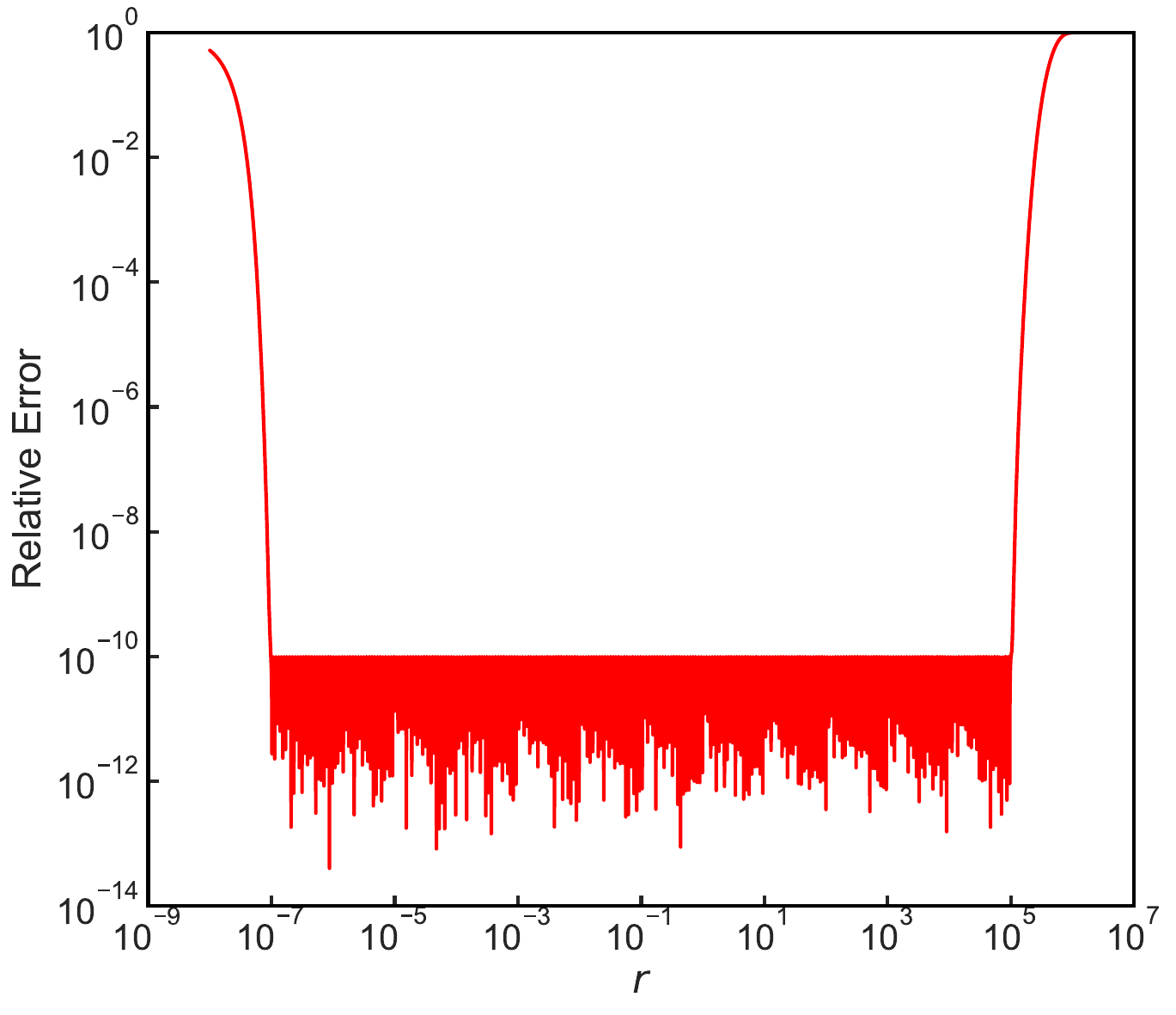}
    \caption{The relative error distribution of SOG approximation of the Coulomb kernel $1/r$ on $[10^{-7},10^{5}]$ by the WBT method.}
    \label{fig:inverse_longrange}
\end{figure}

Finally, we study the performance of the WBT on interval where the end point is close to a singular point. We consider the inverse function with $\alpha=0.5$ over the interval $[10^{-14},10^{10}]$. This is equivalent to an SOG approximation of the Coulomb kernel on $[10^{-7},10^{5}]$ by a simple variable substitution. This example is from Beylkin {\it et al.} \cite{beylkin2019adaptive},  where $10^{-10}$ accuracy in relative error is achieved for the SOG approximation with $P=8$ to represent the long-range part of the BSA. Similarly, we begin with a preliminary SOE approximation using the BSA expansion with indices $n$ from $-203$ to $86$. Since smaller indices exhibit long-range characteristics, we apply the WBT with weight function $w(r)=1/\sqrt{r+10^{9}}$ to reduce terms with indices from $-203 $ to $-52$ while preserving the remaining terms. This approximation achieves $10^{-10}$ relative accuracy across $[10^{-14},10^{10}]$ with only $5$ long-range SOE terms. For purpose of comparison, we present the equivalent SOG approximation as follows,
\begin{equation}
\label{eq::longrange_1/r}
\bigg|\frac{1}{r} - S(r) - \frac{2\sigma\log b}{\Gamma(\frac{1}{2})} \sum_{n= -51}^{86} b^ne^{-b^{2n}\sigma^2r^2} \bigg| \leq \frac{\epsilon}{r},
\end{equation}
where $b =$ 1.22749083347315613, $\sigma =$0.908024474\\99108738 and $\epsilon = 10^{-10}$. The reduced long-range term $S(r)$ reads
\begin{equation}
    \label{eq::reduced}
    S(r)=\sum_{n=1}^{5} w_n e^{-s_n r^2},
\end{equation}
where the weights and exponents are listed in Table~\ref{tab:sn_wn}. Fig.~\ref{fig:inverse_longrange} presents the error over the space, where one can clearly observe a uniform distribution by the WBT. This result demonstrates the advantage of the WBT method in reducing the long-range component of the kernel function.

\begin{table*}[ht!]
\centering
\caption{ Exponents and weights of the $5$ long-range SOE terms in Eq.~\eqref{eq::longrange_1/r}.}
\label{tab:sn_wn}
\begin{tabular}{|c|c|} 
\hline
\(s_n\) & \(w_n\) \\
\hline
\(4.551547331769476\times10^{-10}\) & \(4.955235574250308\times10^{-6}\) \\
\hline
\(2.967225833661697\times10^{-10}\) & \(4.503807233967136\times10^{-6}\) \\
\hline
\(1.69007319959171\times10^{-10}\)  & \(5.145266724826999\times10^{-6}\) \\
\hline
\(6.564737578696118\times10^{-11}\) & \(5.849337004446485\times10^{-6}\) \\
\hline
\(7.549432548035814\times10^{-12}\) & \(6.175199783823309\times10^{-6}\) \\

\hline
\end{tabular}
\end{table*}

\section{Conclusion}
\label{sec::conclusion}
In summary, we propose a novel weighted balanced truncation method for approximating general kernel functions with exponentials. The WBT method incorporates a weight function into the balanced truncation method, resulting in a more accurate approximating precision across the target interval.
Numerical examples demonstrate that the WBT method achieves significant  improvement in accuracy compared to classical model reduction method. As a general approximation technique for kernel functions, it provides effective approximation results for important kernels like the Coulomb interaction. Meanwhile, the WBT method maintains stable performance when handling functions with complex properties, leading to a broad application prospect in physics and scientific computing.

Besides treated as a kernel-independent approximation technique, the WBT can also be regarded as an improved model order reduction method with even broader applicability. Future work will focus on designing efficient applications of the WBT method in cutting-edge fields, such as machine learning and materials computation.


\section*{Acknowledgement}

This work is supported by Natural Science Foundation of China (grants No. 12325113 and 12426304) and SJTU Kunpeng \& Ascend Center of Excellence.

\section*{Conflict of interest}

The authors declare that they have no conflict of interest.


\section*{Data Availability Statement}

The software package generating the data in this work is developed based on the MATLAB and the Multiprecision Computing Toolbox. 
The source code is available at \href{https://github.com/linyuanshen114/VP-WBT}{https://github.com/linyuanshen114/VP-WBT}. 
\\[2ex]


%

\end{document}